\documentclass[english,twocolumn, prl, superscriptaddress, longbibliography]{revtex4-1}
\usepackage[T1]{fontenc}
\usepackage[latin9]{inputenc}
\usepackage{color}
\usepackage{amsmath}
\usepackage{graphicx}
\usepackage{dsfont}
\usepackage[colorlinks=true,linkcolor=blue,urlcolor=blue,citecolor=blue,anchorcolor=blue]{hyperref}
\makeatletter
\usepackage{amsmath}
\usepackage{subfigure}
\usepackage{graphicx, mathtools}
\usepackage[normalem]{ulem}

\makeatother

\usepackage{babel}

\begin{document}

\newcommand{\be}{\begin{equation}}
\newcommand{\ee}{\end{equation}}
\newcommand{\bn}{\begin{eqnarray}}
\newcommand{\en}{\end{eqnarray}}
\newcommand{\ii}{\'{\i}}
\newcommand{\ca}{\c c\~a}
\newcommand{\uc}{\uppercase}
\newcommand{\tb}{\textbf}
\newcommand{\bw}{\begin{widetext}}
\newcommand{\ew}{\end{widetext}}

\title{Thermal Transport across a Continuous Metal-Insulator Transition}

\author{P. Haldar}\email{prosenjit@imsc.res.in}

\author{M. S. Laad}\email{mslaad@imsc.res.in}

\author{S. R. Hassan}\email{shassan@imsc.res.in}

\affiliation{Institute of Mathematical Sciences, Taramani, Chennai 600113, India}

\affiliation{Homi Bhabha National Institute Training School Complex,
Anushakti Nagar, Mumbai 400085, India}

\date{\rm\today}

\begin{abstract}
The celebrated Wiedemann-Franz (WF) law is believed to be robust in metals as long as interactions between electrons preserve 
their  fermion-quasiparticle character.  We study thermal transport and the fate of the WF law close to a continuous 
metal-insulator transition (MIT) in the Falicov-Kimball model (FKM) using cluster-dynamical mean-field theory (CDMFT).  
Surprisingly, as for electrical transport, we find robust and novel quantum critical scaling in thermal transport across the MIT. We unearth the deeper reasons for these novel findings in terms of $(i)$ the specific structure of {\it energy}-current 
correlations for the FKM and $(ii)$ the microscopic electronic processes which facilitate energy transport while simultaneously
 blocking charge transport close to the MIT.  However, within (C)DMFT, we also find that the WF law survives at $T\longrightarrow0$ in the 
incoherent metal right up to the MIT, even in absence of Landau quasiparticles.  
\end{abstract}
   

\maketitle

\section{Introduction}

	In recent years, there have been lots of studies on thermo-electric properties of the strongly correlated materials such as $Bi_2Te_3/Sb_2Te_3$, $LaFe_3CoSb_{12}$ and $CeFe_3CoSb_{12}$ which have myriad applications~\cite{mahan_thermo} in designing new devices. There are different theoretical investigations of the thermoelectric materials~\cite{ziman,kubo1957,mahan}. The Boltzmann theory which is applicable in weakly coupled system where Landau quasi particle picture remains valid. But this theory will not work in the strongly correlated materials as the perturbation theory breaks down in this regime. Kubo formalism has been used in both weak and strong coupling regime. But the drawback of Kubo formalism is that its dynamical nature (frequency dependence) makes difficult to calculate some experimentally accessible quantities such as the thermopower or Seebeck coefficient, Lorenz number and thermal conductivity. Another approach proposed by Shastry~\cite{shastry}, where he formulated the computation of the thermal response to dynamical temperature gradients by neglecting the intricacy of the full dynamics of the Kubo formalism. 

It is interesting to study the effect of disorder in materials along with interaction~\cite{disorderrmp}. Transport properties (electric, thermal) with the controlled disorder can play a vital role in designing new materials. One of the most prominent kind of disorder systems is binary disordered alloy where disorder is induced by creating vacancies in the crystalline order for materials like $A_{x}N$, with A=Ta, Nb,...etc. Here, the disorder strength can be controlled by changing the vacancies of A-atom. 
Falicov-Kimball model (FKM) can describe these materials~\cite{thermal2} well. But the only difference is that FKM accounts for annealed disorder instead of quenched-like disorder in the binary disordered alloy. Therefore, one can investigate the effect of (binary) disorder on the transport properties of these material using FKM within binary alloy analogy~\cite{ourfirstpaper}.   

	Another fascinating features of the thermal transport, in normal metals at low temperature $T$, the celebrated Wiedemann-Franz (WF) law~\cite{wf_law} relates the electrical and thermal 
conductivities via a universal Lorenz number, $L_{0}=\frac{K_{el}(T)}{T\sigma_{xx}(T)}=\frac{\pi^{2}k_{B}^{2}}{3e^{2}}$, the 
Sommerfeld value.  Even in strongly correlated metals~\cite{rareearths}, the WF law still holds as $T\rightarrow 0$ as long 
as the metallic state is a Landau Fermi Liquid (LFL), presumably due to a Ward identity~\cite{castellani}.  Explicit counter-examples are $D=1$ 
Luttinger liquids~\cite{fisher}, cuprates~\cite{htsc} and $f$-electron systems~\cite{steglich} near quantum phase transitions, 
where Landau quasiparticle views break down.  It is well known that Landau quasiparticle picture also naturally breaks down at 
interaction- or disorder-driven metal-insulator transitions (MIT) (at $T=0$).  However, the former are generically first-order, 
and are accompanied by instabilities to more conventional symmetry-broken states at lower $T$, preventing clean study of the
 breakdown of the WF law.  Thus, continuous MITs at $T=0$ turn out to be an ideal playground to study this issue.

    Quite generally, quantum critical fluctuations at a continuous MIT affect critcal features in conductivity.  This has been 
studied in the context of the finite-but low $T$ critical end-point in the $d=\infty$ Hubbard model~\cite{terletska}, and recent 
CDMFT work for the FKM also shows that conductivity~\cite{qcmott} and magneto-transport~\cite{qchall3} exhibit remarkable 
quantum-critical 
scaling at a ``Mott'' QCP.  Whether and how such novel QC features show up in thermal transport is a very interesting, albeit 
scarcely studied, issue.  Thermal transport primarily measures {\it energy} current correlations in solids~\cite{thermal1,thermal2}.  
Most generally, the 
electronic contribution to the thermopower, $S_{el}(T)$, is best interpreted as the entropy of an electric 
current~\cite{springerbook}. 
In weakly correlated metals, $S_{el}(T)\simeq A_{1}T$ is small at 
low $T$.  In strongly correlated Landau Fermi Liquid (LFL) metals, in contrast, $S_{el}(T)=AT$ is sizably enhanced at low $T$,
 passes through a broad 
maximum at intermediate $T$ before asymptoting to the Heikes law~\cite{heikes1961} at high $T>>T_{LFL}\simeq t_{eff}=z_{FL}t$, where $t_{eff}$ is
 the correlation-induced reduction of the bare kinetic energy ($t$) and $z_{FL}$ 
is the Landau quasiparticle residue.  In the Mott insulator, one expects $S_{el}(T\rightarrow 0)$ to diverge owing to the loss 
of carriers upon gap opening.  It is then natural to expect that soft quantum-critical 
fluctuations at a QCP associated with a continuous MIT should also reflect in thermal transport. Moreover, such studies also permit one to analyze Thomson effects using the Kelvin relations~\cite{springerbook}.  In fact, the Thomson co-efficient, which is just heat per unit current and unit temperature gradient, is simply related to the thermopower via $\tau_{th}(T)=T(dS_{el}(T)/dT)$: thus, this quantifies the "specific heat of electricity"~\cite{springerbook}.  That the $\gamma$-co-efficient of the usual constant-volume specific heat diverges at a continuous MIT is well known.  Does the ``specific heat of electricity'' also show a critical divergence at such a MIT?        

	Motivated hereby, we study thermal transport in the simplest lattice model of interacting fermions, the Falicov-Kimball 
model (FKM) in detail within a two-site cluster-DMFT~\cite{ourfirstpaper} within the alloy-analogy formalism.  Specifically, we $(i)$ unearth quantum-critical scaling in thermal transport and correlate it with electrical 
transport, and $(ii)$ examine the microscopic origin of the electronic processes involving energy current which distinguish 
thermal from electrical transport.  The FKM is ideal since it shows a continuous "Mott" MIT within both DMFT~\cite{freericks}, CDMFT~\cite{ourfirstpaper} and also within Coherent Potential Approximation (CPA) with Dynamical Cluster Approximation (DCA)~\cite{jarrell}.  We focus on quantum critical features in thermal transport in the strong-scattering 
regime where 
$k_{F}l\simeq 1$ invalidates quasiclassical Boltzmann approaches, since the very 
concept of well-defined LFL quasiparticles breaks down.    
  
	FKM can be solved exactly within DMFT~\citep{freericks} in infinite dimensional systems. As an advanced mean field type technique DMFT is more reliable method for studying materials properties in three or higher dimension~\cite{RMP1996}. As for the conductivity tensor~\cite{qcmott,qchall3}, it turns out that thermal transport co-efficients can be precisely evaluated
 within our two-site CDMFT~\cite{ourfirstpaper}.  This is because the irreducible cluster resolved particle-hole vertex 
corrections rigorously drop out from the Bethe-Salpeter equations (BSE) for all current-current correlation 
functions~\cite{kotliar}. Further, having explicit closed-form analytical expressions for the cluster propagators,
 $G({\bf K},\omega)$, minimizes the computational cost, even within CDMFT. 

	The plan of the paper is as follows: In Sec. II we describe in details of our model within cluster-DMFT formalism and the calculation of thermal transport (thermopower, thermal conductivity, Lorentz number and Thomson coefficient) using Cluster DMFT formalism. In Sec. III we report numerical result for  the thermal transport using CDMFT and Quantum criticality of thermal transport across the MIT . In Sec. IV we compare our result the with single site DMFT result. We present discussion and conclusions in Sec. V. \\

\section{General formulation for thermal transport within cluster DMFT}

	The Hamiltonian for spinless FKM~\cite{freericks} or equivalent binary-alloy disorder model is
\begin{equation}
H_{FK} = -t\sum_{\langle i,j\rangle} (c^{\dag}_i c_j + h.c.) + U \sum_i x_i c^{\dag}_i c_i + \mu \sum_i  c^{\dag}_i c_i
\end{equation}
on a Bethe lattice with semicircular band density of states (DOS) as an approximation to a three dimensional lattice. Where, $c^{\dag}_i (c_i)$ is the electron creation (annihilation) operator for spinless electron at site i , $x_i$ is variable that can take either 0 or 1 value, $v_i=Ux_i$ is viewed as a static disorder potential for the c-fermions.\\
In our recent work~\cite{ourfirstpaper} we use our recent exact-to-O(1/D) extension of DMFT to solve FKM applying equation of motion. The local Green's function in two-site cluster DMFT is,
\[ \hat{\mathbf{G}}= \left(  \begin{array}{cc}
G_{00}(\omega) & G_{\alpha 0}(\omega) \\
G_{\alpha 0}(\omega) & G_{00}(\omega) \end{array} \right) \] 
where, the matrix element $G_{ij}(\omega)$
\bw
\begin{eqnarray}
\label{eq:11}
G_{ij}(\omega) = \left[\frac{1-\langle x_0\rangle-\langle x_{\alpha}\rangle+\langle x_{0\alpha}\rangle}{\xi_2(\omega)}+\frac{\langle x_0\rangle-\langle x_{0\alpha}\rangle}{\xi_2(\omega)-U} \right] \left[  \delta_{ij} - \frac{F_2(\omega)}{(t-\Delta_{\alpha 0}(\omega))}(1-\delta_{ij}) \right] \nonumber\\
+\left[\frac{\langle x_{\alpha}\rangle-\langle x_{0\alpha}\rangle}{\xi_1(\omega)}+\frac{\langle x_{0\alpha}\rangle}{\xi_1(\omega)-U} \right] \left[  \delta_{ij} - \frac{F_1(\omega)}{(t-\Delta_{\alpha 0}(\omega))}(1-\delta_{ij}) \right]
\end{eqnarray}
\ew
where the bath function $\hat{\Delta}(\omega)$ is related with the local Green's function through suitable self-consistency condition. The self energy is given as,
\begin{equation}
\hat{\Sigma}(\omega) = \hat{\mathcal{G}}^{-1}_0(\omega) - \hat{G}^{-1}(\omega)
\end{equation} 
with $\hat{\mathcal{G}}_0(\omega)$ is the Wiess Green's function, $\hat{\mathcal{G}}_0(\omega)=(\omega + \mu) \mathds{1} - \hat{\Delta}(\omega)$.  
We use the algorithm described in paper~\cite{ourfirstpaper} to find the local Green's function and self energy. In symmetric basis (cluster momentum basis), we can write $G_{S}=(G_{00} + G_{\alpha 0})$ and $G_{P}=(G_{00} - G_{\alpha 0})$ with S=(0,0,...) and P=($\pi,\pi,..$).

Now, using Kubo-Greenwood formula we calculate the transport properties. In contrast to the electrical conductivity which involves the particle current, 
${\bf j}_{e}=\sum_{\bf q}{\bf v}_{\bf q}c_{\bf q}^{\dag}c_{\bf q}$ with ${\bf v}_{\bf q}=\nabla_{\bf q}\epsilon_{\bf q}$ for an
 unperturbed band structure $\epsilon_{\bf q}$, the heat current required for thermal transport is more 
complicated~\cite{freericks}

\be
{\bf j}_{\bf Q}=\sum_{\bf q}(\epsilon_{q}-\mu){\bf v}_{\bf q}c_{\bf q}^{\dag}c_{\bf q}  + \frac{U}{2}\sum_{\bf q,q'}W({\bf q}-{\bf q'})({\bf v}_{\bf q}+{\bf v}_{\bf q'})c_{\bf q}^{\dag}c_{\bf q'}
\ee
with $W({\bf q})=\frac{1}{N}\sum_{j}e^{-i{\bf q}.{\bf R}_{j}}d_{j}^{\dag}d_{j}$, and thus the heat current contains both 
``kinetic'' and ``potential'' terms.  Quite generally, in terms of the Onsager co-efficients,
$L_{lm} (l,m=1,2)$ with $L_{12}=L_{21}$, one finds

\be
\sigma_{dc}(T)=e^{2}L_{11}
\ee

\be
S_{el}(T)=-\frac{k_{B}}{e} \frac{L_{12}}{L_{11}}
\ee
and

\be
K_{el}(T)=\frac{k_{B}^{2}}{T} \frac{L_{11}L_{22}-L_{12}^{2}}{L_{11}}
\ee

   The $L_{lm}$ can themselves be expressed in terms of the cluster propagators by noticing that these are the zero-frequency 
limit of the analytically continued ``polarization'' operators.  Explicitly,
$L_{lm}=\lim_{\omega\rightarrow 0} Re{\frac{iL_{lm}(\omega)}{\omega}}$, with

\be
L_{11}(i\omega_{n})=\int_{0}^{\beta} d\tau e^{i\omega_{n}\tau} Tr\frac{\langle T_{\tau}e^{-\beta H}j_{e}(\tau)j_{e}(0)\rangle}{Z}
\ee
\be
L_{12}(i\omega_{n})=L_{21}(i\omega_{n})=\int_{0}^{\beta} d\tau e^{i\omega_{n}\tau} Tr\frac{\langle T_{\tau}e^{-\beta H}j_{e}(\tau)j_{Q}(0)\rangle}{Z}
\ee
and

\be
L_{22}(i\omega_{n})=\int_{0}^{\beta} d\tau e^{i\omega_{n}\tau} Tr\frac{\langle T_{\tau}e^{-\beta H}j_{Q}(\tau)j_{Q}(0)\rangle}{Z}
\ee

  In absence of vertex corrections to transport co-efficients, the $L_{lm}$ can finally be expressed in terms of the cluster
 propagators, $G({\bf K},\omega)$.  $L_{11}$ is the same as the one derived for the $dc$ conductivity $\sigma_{xx}(T)$ 
earlier~\cite{qcmott}:
\begin{equation}
L_{11} = \sum_{a=S,P} \frac{T\sigma_{0}}{e^2}\int d\epsilon  \rho_{a}(\epsilon) \int d\omega (\frac{-df(\omega)}{d\omega}) A_{a}^{2}(\epsilon,\omega)
\end{equation}
The Onsager co-efficient relevant for heat transport is most conveniently given in the two-site cluster
 bonding-anti-bonding basis ($S,P$ channels~\cite{ourfirstpaper}) as the sum of the ``kinetic'' and ``potential'' contributions
 as sketched above, $L_{12}=L_{12}^{k}+L_{12}^{p}$. and following Freericks {\it et al.}~\cite{thermal1} for our two-site CDMFT, 
this reads

\be
L_{12}=\sum_{a=S,P}\frac{T\sigma_{0}}{e^{2}}\int d\epsilon \rho_{a}(\epsilon)\int d\omega (-\frac{df(\omega)}{d\omega})\omega A_{a}^{2}(\epsilon,\omega)
\ee
 while $L_{22}$ is given by

\be
L_{22}=\sum_{a=S,P}\frac{T\sigma_{0}}{e^{2}}\int d\epsilon \rho_{a}(\epsilon)\int d\omega (-\frac{df(\omega)}{d\omega})\omega^{2}A_{a}^{2}(\epsilon,\omega)
\ee  
  As for the conductivity tensor~\cite{qchall3}, it turns out that thermal transport co-efficients can be precisely evaluated
 within our two-site CDMFT~\cite{ourfirstpaper}.  
\begin{figure*}
\includegraphics[width=1.5\columnwidth , height= 
1.\columnwidth]{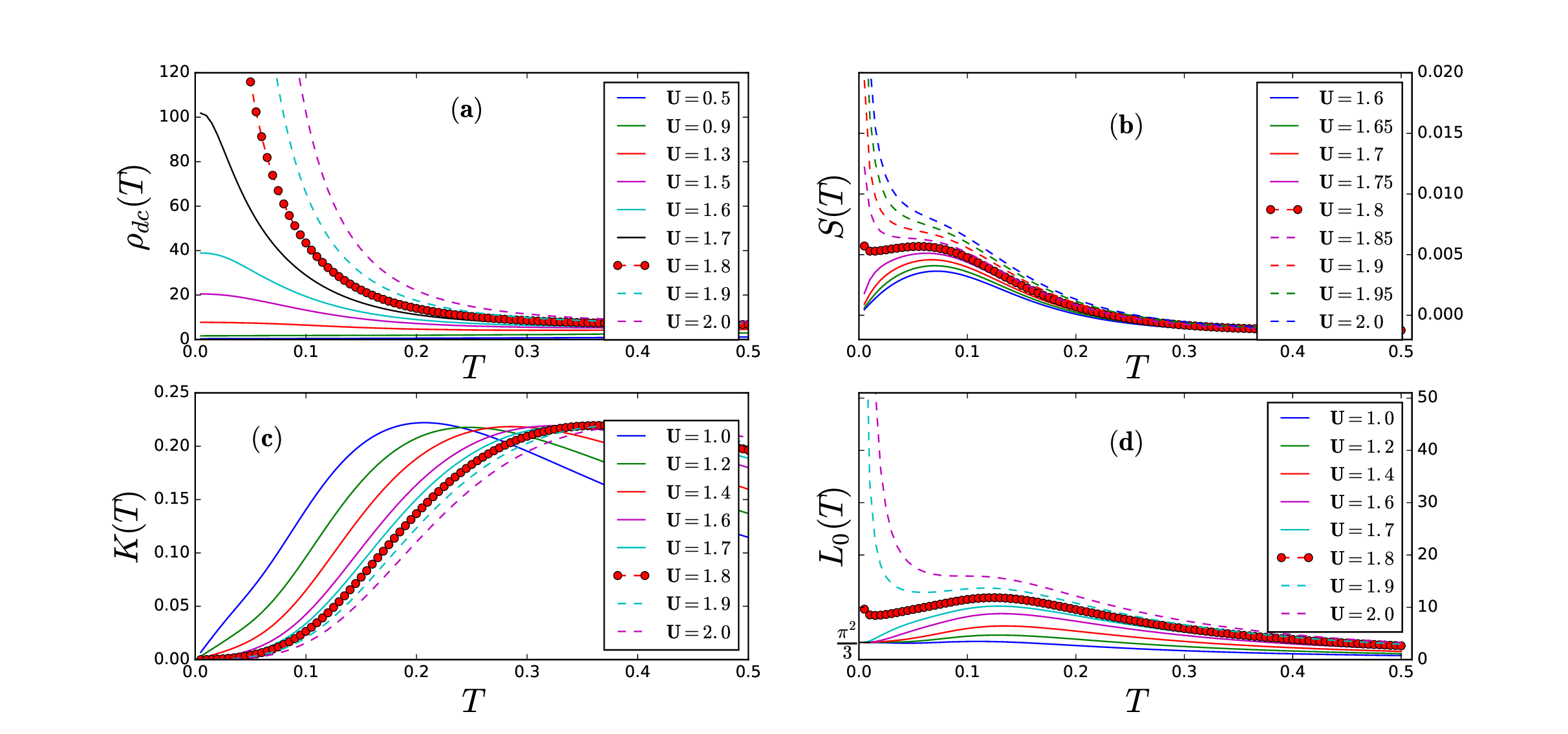} 
\caption{(Color online) $dc$ resistivity $\rho_{dc}(T)$ ($a$), thermopower $S_{el}(T)$ ($b$), thermal conductivity $K_{el}(T)$ ($c$) and Lorenz number $L_{0}(T)$ ($d$) for the FKM as functions of $U/t$.  At the Mott QCP (bold red circles) at $(U/t)_{c}=1.8$, both $S_{el}(T), L_{0}(T)$ attain finite values, cleanly separating metallic and insulating behavior.  Concomitantly, $\rho_{dc}(T\rightarrow 0)$ diverges and $K_{el}(T)\simeq T^{1+\nu}$ with $\nu\simeq 4/3$~\cite{qcmott}.}
\label{fig:fig1}
\end{figure*}

\section{Result within Cluster DMFT}

In this section, we show the result of the thermal transport with two site CDMFT approach. For convenience consider non-interacting electrons half-bandwidth as unity i.e. 2t=1. Since we aim to correlate specific features in electrical and thermal transport with each other, we start by recapitulating
 $dc$ resistivity. 

In Fig.~\ref{fig:fig1}(a), we exhibit the $dc$ resistivity, $\rho_{dc}(U,T)$ as a function of $U$ as the system is driven 
through a continuous MIT at $U_{c}=1.8$~\cite{qcmott}.  It is clear that at intermediate $0.95 < U < 1.8$, clear
pseudogap signatures appear in $\rho_{dc}(T)$ over a progressively wider $T$-range, between the high-$T$ incoherent metal 
and a low-$T$ bad metal, before the MIT occurs for $U \geq 1.8$.  This feature is associated with proximity to the 
``Mott'' quantum critical
point (QCP) occuring between a $T=0$ very bad metal and a ``Mott'' insulator at $U_{c}$.  We are interested in how this
 Mott quantum criticality manifests in thermal transport.  

In Fig.~\ref{fig:fig1}(b), we show how the electronic contribution to the thermopower varies across the continuous MIT.  
Several features stand out:
$(i)$ for weak-to-intermediate $U< 0.9$, $S_{el}(T)\simeq AT$ at low $T<0.025t$ is small (not shown), as expected for a weakly 
correlated metal, and goes hand-in-hand with $\rho_{dc}(T)\simeq const$ at low $T$.  $(ii)$ In the intermediate-to-strong 
coupling ($0.9 < U < 1.7$)
 regime, where one is in the increasingly bad-metallic low-$T$ regime, $S_{el}(T)$ is still linear-in-$T$, but is significantly
 enhanced by factors of $O(50-100)$ over its weakly correlated values.  $S_{el}(T)$ also exhibits a broad peak around 
$T^{*}\simeq 0.04t$, before continuously falling off to achieve the Heikes value~\cite{heikes1961,heikes} at very high $T$.  It is very interesting that 
$S_{el}(U,T)=A(U)T$ with $A(U)$ increasing with $U$ holds throughout this very bad metallic regime, even as
 $\rho_{dc}(T\rightarrow 0)\simeq 100\hbar/e^{2}$.  This is the regime in which no quasiclassical 
Boltzmann view of transport is tenable, since application of Drude-Boltzmann ideas would necessarily yield
$k_{F}l<1$ (where no $1/k_{F}l$-expansion is possible).  Since thermopower features result solely from a non-Landau 
quasiparticle cluster propagator within CDMFT, this implies that this low-$T$ enhancement in $S_{el}(T)$ involves non-Landau-FL
 quasiparticle (branch-cut continuum) excitations.  Just before the MIT, $S_{el}(T\rightarrow 0)$ is still linear in $T$, but
 is enhanced by a factor of about 100 relative to its small $U$ value.  $(iii)$ Finally, precisely at the QCP $U=1.8$, 
clear anomalies obtain: $S_{el}(T)$ increases with decreasing $T$ right down to $T\rightarrow 0$, but achieves a {\it finite}
 value.  For $U >1.8$, opening of the ``Mott'' gap in the one-electron density-of-states~\cite{ourfirstpaper} produces a 
divergent $S_{el}(T\rightarrow 0)$.  This is not a violation of the Nernst theorem, since $\rho_{dc}(T\rightarrow 0)$ 
simultaneously diverges.

\begin{figure}
\includegraphics[width=1.\columnwidth , height= 
1.\columnwidth]{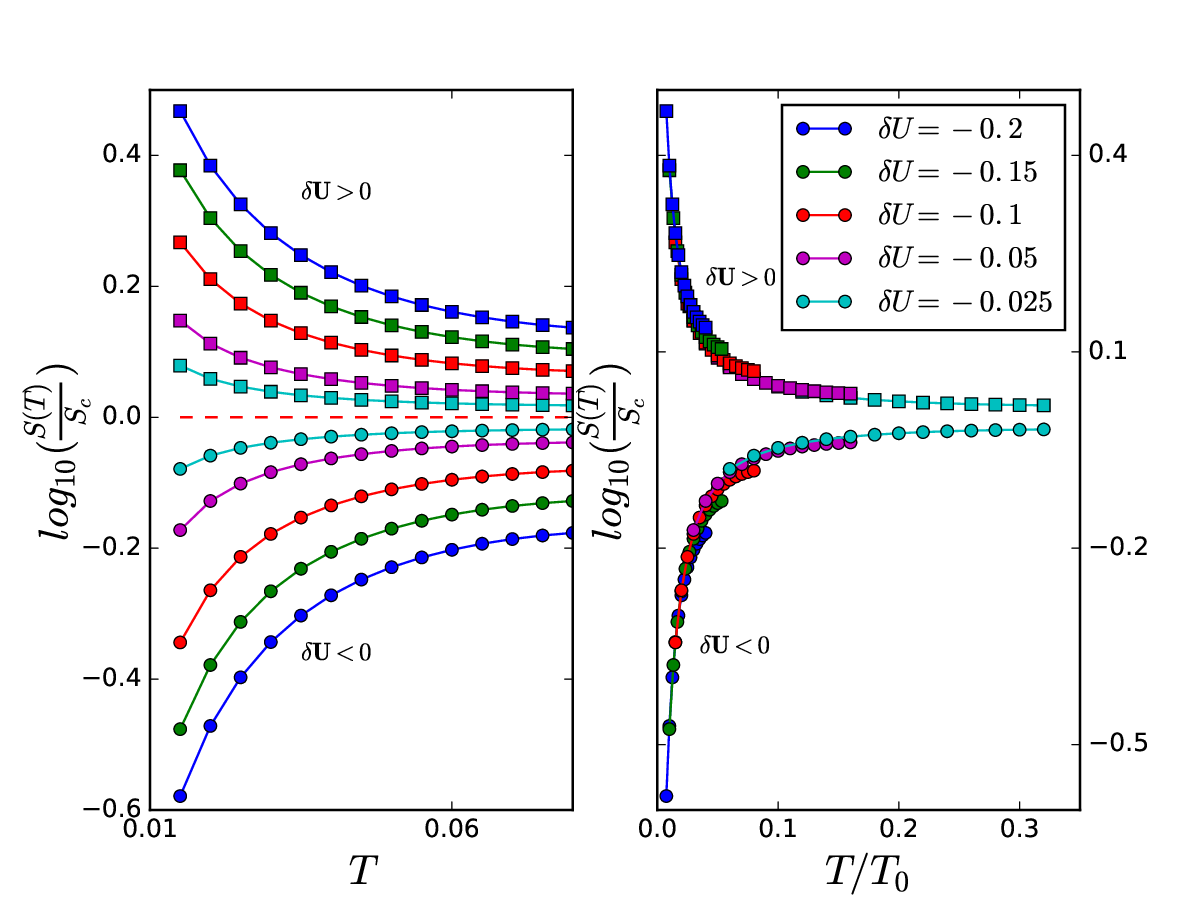} 
\caption{(Color online) Mott Quantum critical scaling in thermopower $S_{el}(U/t,T)$ across the MIT.  Log$(S_{el}(T)/S_{c})$ vs
 $T$ exhibits almost perfect ``mirror symmetry'' around $(U/t)_{c}$ (left panel).  Collapse of metallic and insulating
 curves onto two "universal" curves upon scaling T axis by $T^{th}_{0}$(right panel).  This is evidence that Mott quantum critical scaling in
 electrical transport~\cite{qcmott} extends to thermal transport as well. }
\label{fig:fig2a}
\end{figure}

\begin{figure}
\includegraphics[width=1.\columnwidth , height= 
1.\columnwidth]{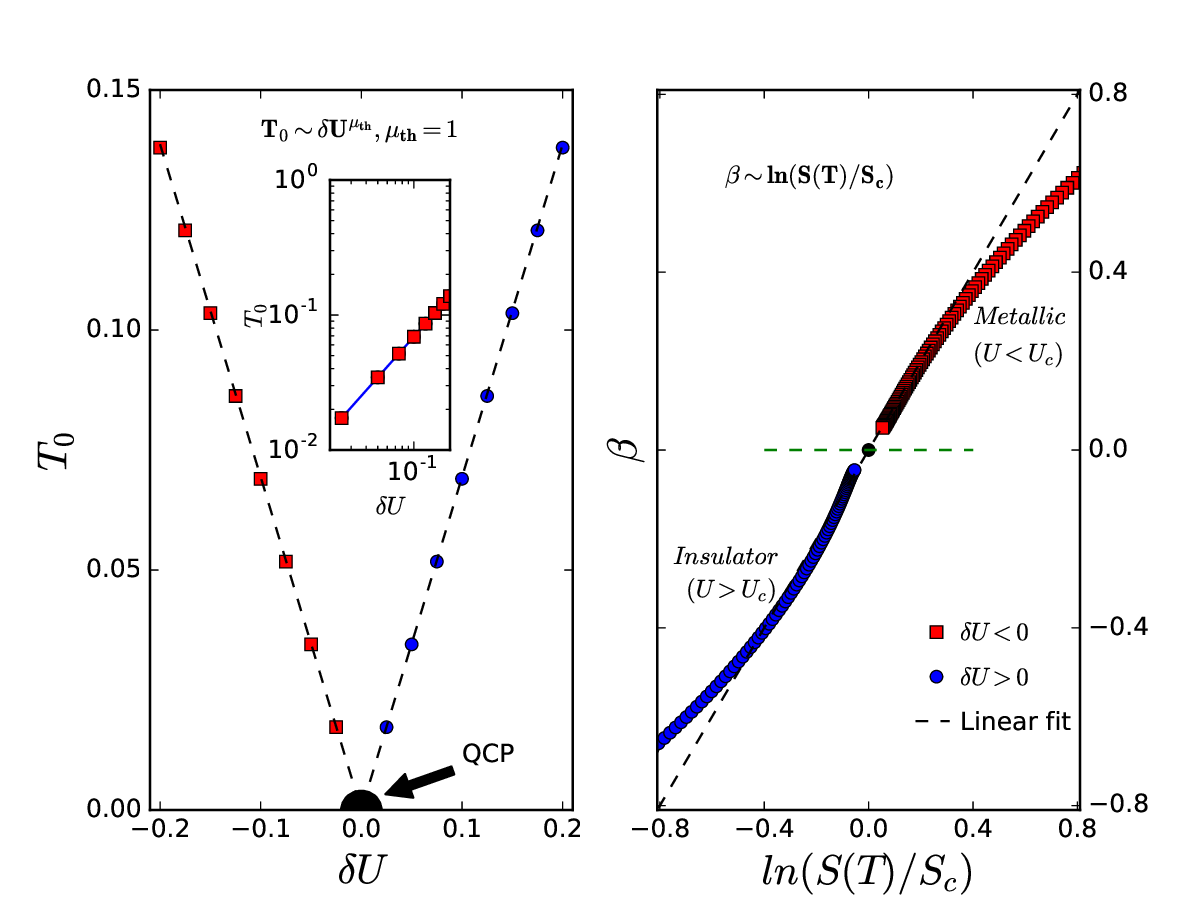} 
\caption{(Color online)  $T_{0}^{th}(\delta U)=c|\delta U|^{\mu_{th}}$ with 
$\mu_{th}=1$ (left panel).  The ``beta function'' varies like $\beta(s)\simeq$ log$(s)$ with $s=S_{el}(T)/S_{c}$ close 
to the MIT and is continuous across $U_{c}$ (right panel) }
\label{fig:fig2b}
\end{figure}

  It is clear from Fig.~\ref{fig:fig1}(b) that $S_{el}(U,T\rightarrow 0)$ curves fan out to either metallic or insulating 
values, except at the ``Mott'' QCP, where $S_{el}$ is finite.  This suggests that, like electrical transport~\cite{qcmott}, 
thermal transport should also exhibit characteristic quantum critical features.  To unravel this novel possibility, we repeat 
earlier procedure~\cite{qcmott} for 
thermopower by making the metallic and insulating curves fall on to two ``universal'' curves by scaling both with a 
$U$-dependent scale, $T_{0}^{th}(U)$. 
In the left panel of Fig.~\ref{fig:fig2a}, we exhibit log$(S_{el}(T)/S_{el}^{(c)})$ versus $T$.  Remarkably, this bares clear signatures of 
``mirror'' symmetry, exactly as in electrical transport.  This strongly 
presages novel ``Mott'' quantum critical features in thermal transport as well.  More clinching support for such criticality
 is seen in right panel of Fig.~\ref{fig:fig2a}, where we show
log$(S_{el}(T)/S_{el}^{(c)})$ versus $T/T_{0}^{th}(U)$ as done earlier~\cite{qcmott}.  Remarkably, we find $(i)$ clear
 ``mirror'' symmetry between metallic and insulating curves around the critical $S_{el}(U_{c})$, 
and $(ii)$ $T_{0}^{th}(\delta U)=c_{th}|\delta U|^{\eta}$ with $\eta=1$ (in Fig.~\ref{fig:fig2b} left panel). 
To further cement this unusual idea, we also show in the right panel of Fig.~\ref{fig:fig2b} 
the "beta"-function (or the Gell-Mann Low function) for thermopower, 
$\beta_{th}(s)=d[$log$(s)]/d[$log$(T)]$ versus $s$, with $s=(S_{el}(T)/S_{c}(T))$ and $S_{c}(T)$ being the critical 
thermopower right at the MIT (red circled curve in Fig.~\ref{fig:fig1}(b)).  
Remarkably, we find $\beta_{th}(s)\simeq$ log$(s)$ near the MIT, exactly as found before for the $dc$ conductivity. 
 This conclusively establishes novel quantum-critical scaling of the thermopower at the 
``strong localization'' MIT as well.

   Appearance of such quantum-critical scaling in thermopower at the MIT is very surprising, and calls
for deeper analysis.  Since $S_{el}(T)$ measures ``mixed'' electrical current-energy current correlations, these
features must originate from long-time behavior of $\langle j_{e}(\tau)j_{Q}(0)\rangle$.  Let us look more closely at this 
term.  The energy current, in contrast to the electrical current, involves {\it three} sites, and reads~\cite{prelovsek}

\be
j_{i,Q}=t^{2}(ic_{i-\delta}^{\dag}c_{i+\delta} + h.c) -\frac{U}{2}(j_{i-\delta,i}+j_{i,i+\delta})(n_{i,d}-\frac{1}{2})
\ee
where we have relabelled $c\rightarrow c_{\uparrow},d\rightarrow c_{\downarrow}$, $\delta$ denotes nearest neighbors of site$i$,
 and $j_{i,i+\delta}$ is the electrical current operator.  For the FKM, we have 
$[n_{i,d},H]=0$ for all $i$, and thus $n_{i,d}=0,1$ only.  The expression for $j_{i,Q}$ now simplifies to a revealing form

\be
j_{i,Q}= t^{2}(ic_{i-\delta}^{\dag}c_{i+\delta} + h.c) \pm\frac{U}{4}(j_{i-\delta,i}+j_{i,i+\delta})
\ee
for $(+,-)$ corresponding to $n_{i,d}=0,1$.  Thus, for the FKM, we find that $j_{i,Q}$ is directly related to the electrical 
current operator, providing direct insight into the underlying reason for emergence of very similar quantum critical scaling 
responses in $\rho_{dc}(T)$~\cite{qcmott} and $S_{el}(T)$ above.  Simply put, energy current correlations mirror those of the 
electrical current.

   Armed with these positive features, we now study the electronic contribution to the thermal conductivity, $K_{el}(T)$, in 
Fig.~\ref{fig:fig1}(c).  In the small $U$ regime, $K_{el}(T)\simeq A_{2}T$ is linear in $T$, as would be expected for a weakly 
\begin{figure*}
\includegraphics[width=1.5\columnwidth , height= 
1.\columnwidth]{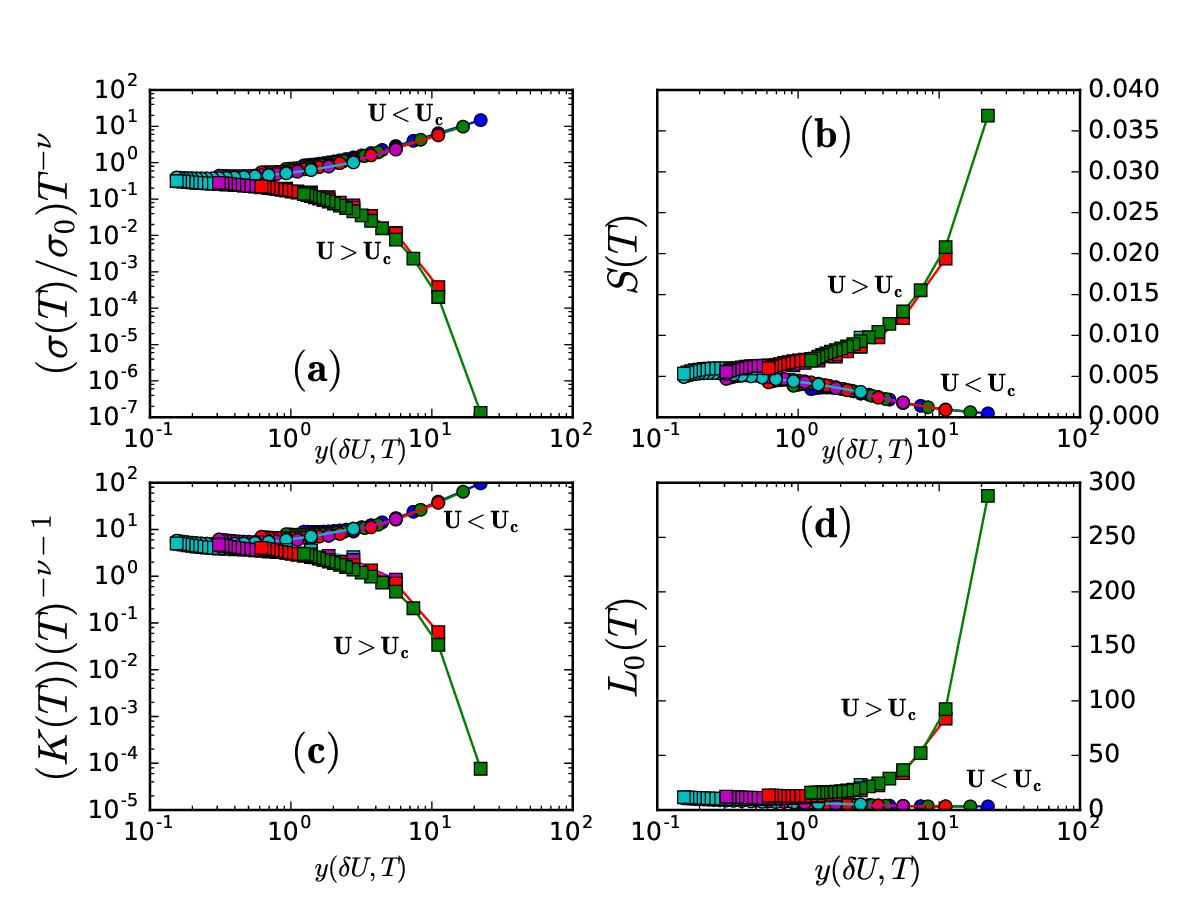} 
\caption{(Color online) Quantum critical scaling in scaled electrical
 conductivity $T^{-\nu}\sigma_{dc}(T)$ (panel ($a$)), thermopower $S_{el}(T)$ (panel ($b$)), scaled thermal conductivity, $T^{-1-\nu}K_{el}(T)(panel ($c$))$ and Lorenz number (panel ($d$)) when plotted as 
functions of the ``scaling variable'' $y(U,T)=|U-U_{c}|/U_{c}T$, demonstrating clean quantum critical scaling in electrical 
as well as thermal transport at the Mott QCP.}
\label{fig:fig3}
\end{figure*}
correlated metal, with transport being determined by a LFL.  This is the regime where 
$\rho_{dc}(T\rightarrow 0)\simeq const$, and formally corresponds to the weak scattering regime where $k_{F}l>>1$ 
holds (this is thus the regime where self-consistent Born approximation (SCBA) applies).  As we enter the intermediate-to-strong
 scattering regime with $0.95 < U < 1.8$, progressive bad metallicity in resistivity
goes hand-in-hand with emergence of a low-energy scale in $K_{el}(T)$, where its power-law-in-$T$ ($K_{el}(T)\simeq T^{n}, n>1$)
 behaviour at intermediate-$T$ crosses over to a linear-in-$T$ variation as $T\rightarrow 0$.  Precisely at $U_{c}=1.8$, we 
find $K_{el}(T)\simeq T^{1+\nu}$.  This behavior is characteristic of heat conductivity arising from non-fermionic excitations. 
 In our case, such collective modes can only be of electronic origin: these are the low-energy particle-hole fluctuations, which
 remain low-energy excitations in the insulator when charge degrees of freedom are frozen out at low energies.  
Upon closer inspection, we see that the linear-in-$T$ contribution gives way to a power-law behavior
($K_{el}(T)\simeq T^{1+\nu}, 0<\nu<1$) right down to $T=0$ for $U=1.8$ within our numeric, {\it precisely} where the MIT 
occurs.  This finding is completely consistent with breakdown of the LFL quasiparticle description in the quantum critical 
region associated with the MIT.
\begin{figure}
\includegraphics[width=1.\columnwidth , height= 
1.\columnwidth]{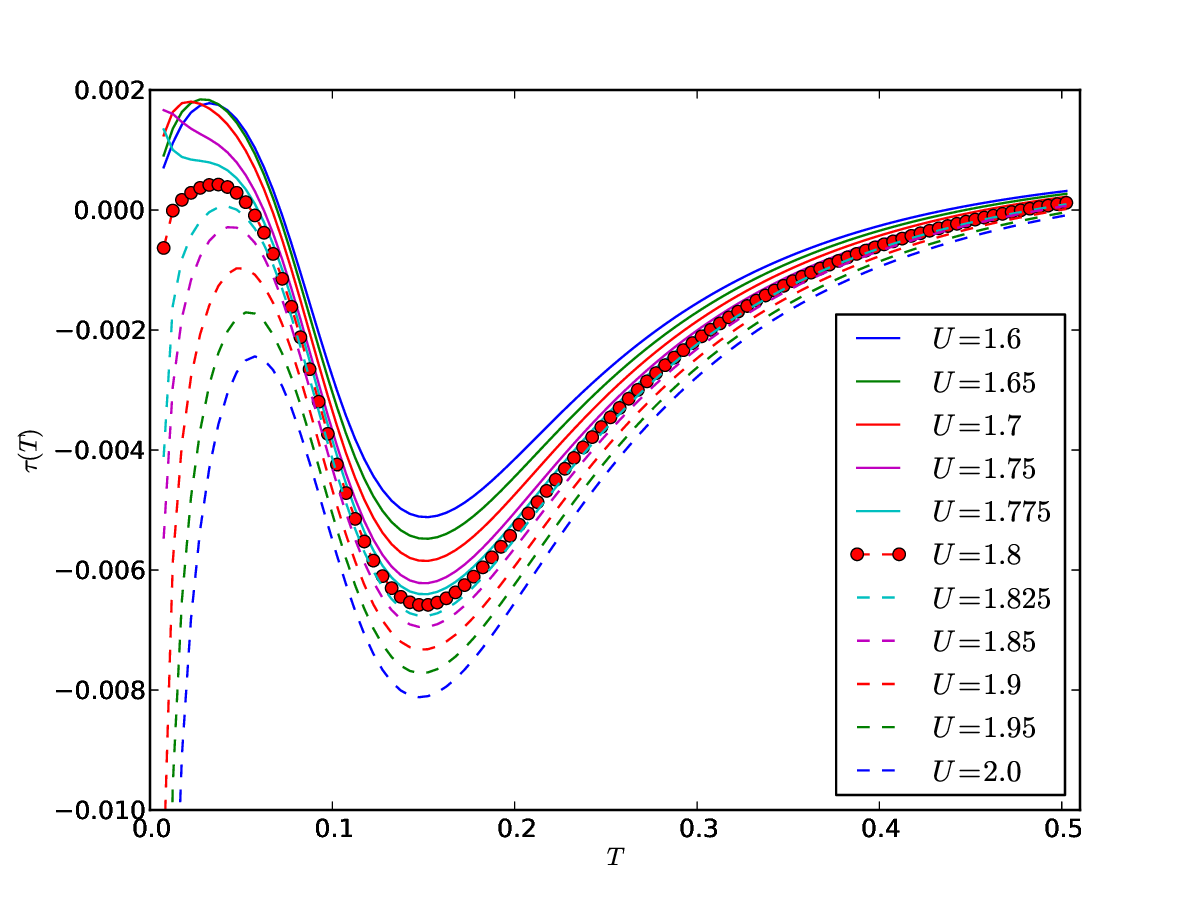} 
\caption{(Color online) Thomson Co-efficient $\tau_{th}(T)$ for FKM as a function of $U/t$}
\label{fig:fig4}
\end{figure}

   Even more insight into the breakdown of the LFL quasiparticle description close to the MIT is provided by examination of 
the $T$-dependent Lorenz number, $L_0 (T)=K_{el}(T)/T\sigma_{xx}(T)$, as a function of 
$U$.  In Fig.~\ref{fig:fig1}(d), we exhibit $L_0 (U,T)$ across the MIT.  Throughout the metallic phase, including the 
very bad metal,
 $L_0 (T\rightarrow 0)=\frac{\pi^{2}}{3}$ (in units of $k_{B}=1=e$), even though 
$L_0(T)$ exhibits significant $T$-dependence up to the lowest $T$, especially for $U>1.4$, implying no breakdown of the WF law 
in the metallic phase.  Precisely at the MIT, however, 
$L_0(T\rightarrow 0)\simeq 10$, indicating breakdown of the WF law exactly at the MIT.  In the insulator ($U >1.8$), 
$L_0(T\rightarrow 0)$ diverges, as it must, since $K_{el}(T)\simeq T^{3}$ while 
$\rho_{dc}(T)\simeq$ exp$(E_{g}/k_{B}T)$.  Our finding is remarkable because, whilst the resistivity shows clear precursor
 features of impending proximity to the MIT via progressive enhancement of 
bad-insulating and very bad metallic regimes beginning from $U =0.95$, both $S_{el}(T)$ and $K_{el}(T)$ continue to display
 apparently conventional behavior right up to the MIT.  Further, spectral responses clearly show non-Landau-FL metallicity~\cite{ourfirstpaper}, and while one may argue for a non-WF behavior at any $T\neq 0$, our results indicate no breakdown of the WF law 
at $T=0$.  
\begin{figure*}
\includegraphics[width=1.5\columnwidth , height= 
1.\columnwidth]{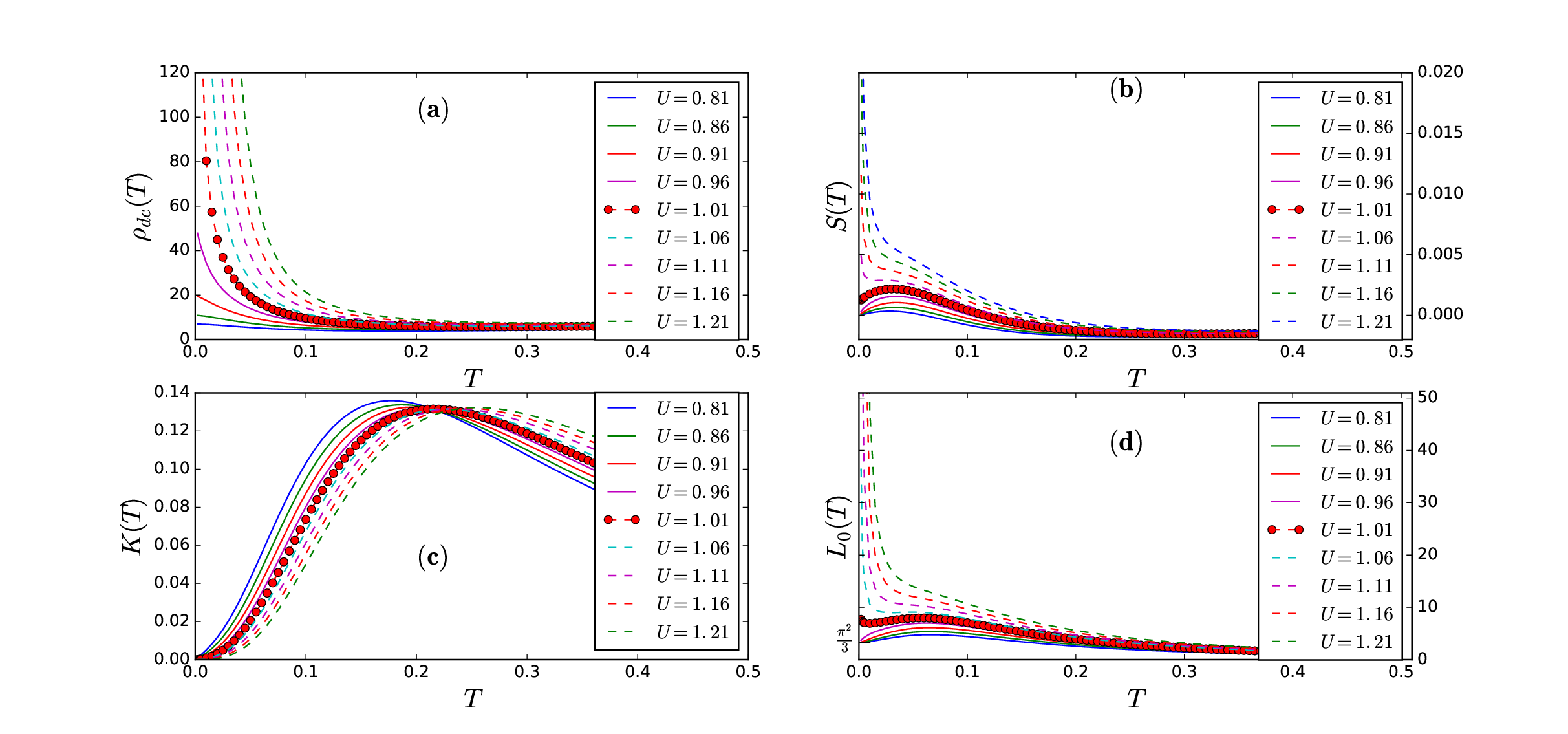} 
\caption{(Color online) Results similar to those in Fig.~\ref{fig:fig1}, but now using single-site DMFT.}
\label{fig:fig5}
\end{figure*}

   Remarkably, upon proper rescaling, it now turns out that $\sigma_{xx}(T), S_{el}(T), K_{el}(T)$ and $L_0(T)$ all exhibit clear 
quantum-critical scaling features.  At the QCP, we find (not shown) that $K_{el}(T)\simeq T^{7/3}=T^{1+\nu}$ with $\nu=4/3$.
Recalling that $\nu=4/3$ is precisely the correlation length exponent we find for the $dc$ conductivity~\cite{qcmott}, this 
suggests an alternative way to exhibit quantum critical scaling that bares the link between electrical and thermal transport. 
   
In Fig.~\ref{fig:fig3}, we find that $T^{-4/3}\sigma_{xx}(T)/\sigma_{0}, S_{el}(T),
T^{-7/3}K_{el}(T)$ and $L_0 (T)$ exhibit clear collapse of the metallic and insulating curves onto two clear branches when plotted as
 a function of the "scaling variable" $y=|U-U_{c}|/U_{c}T$, $i.e$, as a function of the distance from the ``Mott'' QCP.  
Since $\sigma_{xx}(U)\simeq (U_{c}-U)^{4/3}$ as found earlier~\cite{qcmott}, $\nu=4/3$ and $z=1$, as expected for the FKM. 
 Further, $z\nu=4/3 > (2/d)$ implies that the Harris criterion holds, a implying a genuinely {\it clean} QCP.  Again, these
 features reflect the finding above, where energy
current correlations simply mirror the electrical current correlations for the FKM, providing direct microscopic rationale
 for closely related quantum-critical transport in both.  We are aware of only one previous study~\cite{roemer,schreiber} where this 
issue was studied phenomenologically, by using the {\it experimental} conductivity as an input into the Kubo formula for 
the $L_{lm}$.  In contrast, our results emerge from a truly microscopic
CDMFT formulation for the FKM, and our finding of $z=1$ is very different from $z=3$ and $\nu=1$ (latter taken from 
experimental conductivity data).  It is also different from $z=d$ found~\cite{wegner,belitz} for scaling
in the non-interacting disorder model.  Together with Mott-like criticality in transport~\cite{qcmott}, these differences 
reflect the qualitatively distinct ``strong coupling'' nature of the QCP in the FKM.

	Finally, using the Kelvin formula, we now show the Thomson co-efficient as a function of $U/t$ across the MIT. In Fig.~\ref{fig:fig4}, we show $\tau_{th}(U/t,T)$.  In the metallic phase, right up to 
$(U/t)=1.7$, the Thomson co-efficient exhibits a weak $T$-dependence at high $T$, changes sign at a low-to-intermediate 
$T_{1}\simeq O(0.08t)$, passes through a maximum around $0.5T_{1}$ before vanishing 
linearly at lowest $T$.  Exactly at the MIT, qualitative changes occur: $\tau_{th}(U>U_{c},T)$ now exhibits two distinct
regimes where $d\tau_{th}(T)/dT$ changes sign 
(around $1.13t$ and $0.05t$)before asymptoting to a {\it finite} negative value in the insulator.  Remarkably, much alike 
the way in which the $\gamma$-co-efficient of the usual specific heat at constant 
volume diverges upon approach to the MIT, we find that the $\gamma$-co-efficient of the ``specific heat of electricity'',
 defined as $\gamma_{e}=(dS_{el}(T)/dT)$, progressively increases with $U/t$ right 
up to the MIT, diverging at the ``Mott'' QCP.

\section{Single-Site DMFT Results for Thermal Transport}

	Here, we compare our result with single site DMFT~\cite{thermal2} result. For single site DMFT on Bethe lattice local self energy $\Sigma (\omega)$ reads,
\be
\Sigma(\omega) = U \langle x_i\rangle + \frac{U^2 \langle x_i\rangle(1-\langle x_i\rangle)}{\omega - U(1-\langle x_i\rangle)-t^2G_{loc}(\omega)}
\ee
The spectral function, $A(\mathbf{k},\omega)= -\frac{1}{\pi} Im G(\mathbf{k},\omega)$ with $G(\mathbf{k},\omega)^{-1}=\omega -\epsilon_k -\Sigma(\omega)$. Inserting $A(\mathbf{k},\omega)$ in Kubo-Greenwood formula we calculate current-current correlation function~\cite{freericks,thermal1}. It is well known that for single-site DMFT irreducible vertex correction vanishes in the Bethe-Salpeter equation, so only the bare bubble contributes.
\begin{figure}
\includegraphics[width=1.\columnwidth , height= 
1.\columnwidth]{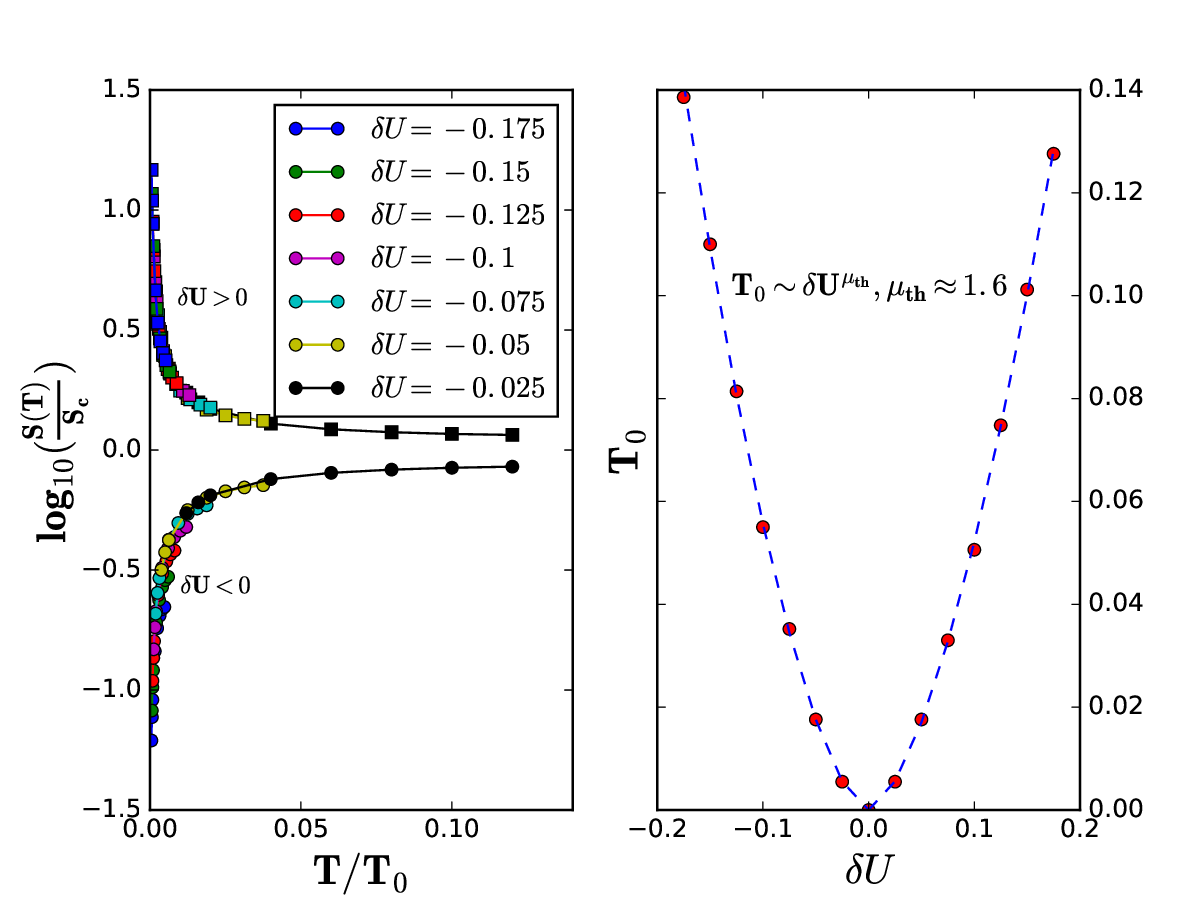} 
\caption{(Color online) (a) In left panel, $log(S_{el}(T)/S_c)$ vs $T/T_0$ and (b) in right panel, $T^{th}_0(\delta U)$ vs $\delta U$ within single-site DMFT. }
\label{fig:fig6}
\end{figure}
\begin{figure*}
\includegraphics[width=1.5\columnwidth , height= 
1.\columnwidth]{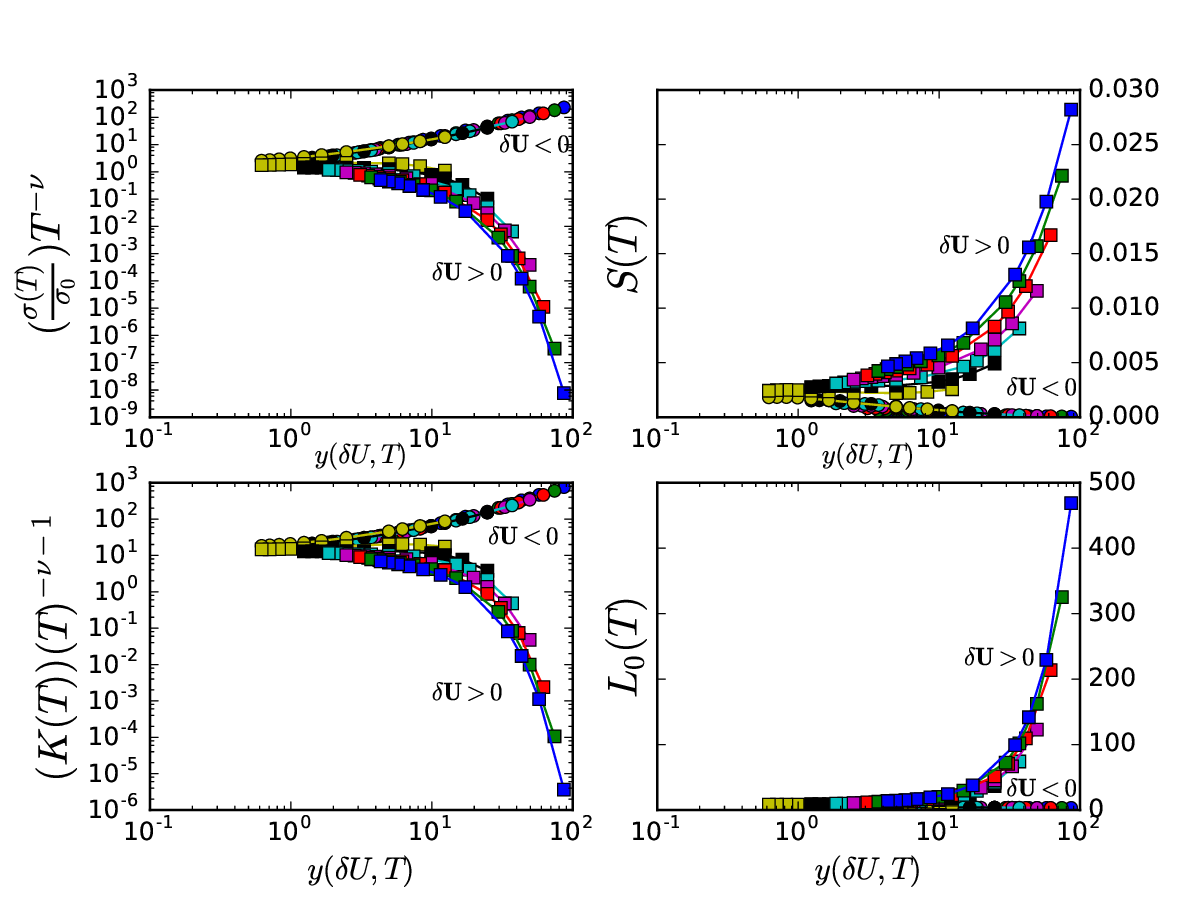} 
\caption{(Color online) Results similar to those obtained in Fig.~\ref{fig:fig3}, but now using single-site DMFT.}
\label{fig:fig7}
\end{figure*}

   We now show single-site DMFT results for electrical and thermal transport.  In $dc$ resistivity across the 
MIT, shown in Fig.~\ref{fig:fig5} (which now occurs at a $(U/t)_{c}^{DMFT}=1.1$), we see features very similar to those found in
 CDMFT.  However, $(i)$ $\rho_{dc}(T)$ at $U_{c}$ now attains values $O(40)\hbar/e^{2}$, much smaller than the 
$O(200)\hbar/e^{2}$ found in CDMFT.  Correspondingly, $S_{el}(T), K_{el}(T)$ and $L_{0}(T)$ exhibit very similar behavior to 
that found in CDMFT, as shown in Fig.~\ref{fig:fig5}.  At first sight, one may thus conclude that no qualitative difference
 exists between DMFT and CDMFT results.

   However, closer inspection of DMFT results, obtained by performing the same scaling analysis as the one done in the main
 text, reveals crucial differences between DMFT and CDMFT results.  Comparing scaling for $S_{el}(T)$ within DMFT in 
Fig.~\ref{fig:fig6} to those obtained from CDMFT in Fig.~\ref{fig:fig2a} and Fig.~\ref{fig:fig2b} in the previous section reveals that
$(i)$ scaling holds over a much narrower window in DMFT compared with CDMFT, and $(ii)$ $\mu_{th}^{DMFT}=1.2$, compared to
 $\mu_{th}=1$ in CDMFT.  It is thus more difficult to discern clean extended scaling behavior from DMFT results, and CDMFT 
clearly performs better in this respect.

   Moreover, repeating the analysis leading to Fig.~\ref{fig:fig3}, we exhibit the results in
 Fig.~\ref{fig:fig7}.  It is now clear that the scaling features in $S_{el}(T), L_{0}(T),T^{-\nu}\sigma_{xx}(T)$ and
$T^{-1-\nu}K_{el}(T)$ are of much poorer quality than those obtained from two-site CDMFT results. 

Comparing with CDMFT results, several features stand out.  These reveal very interesting differences between DMFT and CDMFT results, and we use these to propose that extensions of DMFT to include short-range spatial correlations seem to be {\it necessary} to discuss novel quantum critical scaling in thermal transport at the MIT.

   Thus, while critical features in electrical transport may be adequately captured by single-site DMFT as above (though 
the critical exponents $z$ and $\nu$ are, as expected, different), we find that description of energy transport, and, in 
particular, much better elucidation of
quantum critical thermal transport, requires cluster extensions capable of properly distinguishing between 
non-local aspects entering the distinct microscopic processes which underlie energy transport, as opposed to charge transport. 

\section{Discussion and Conclusion}

What is the microscopic origin of boson-like collective modes that can provide a distinct channel for heat conduction which 
simultaneously blocks charge transport?  It is most instructive to 
invoke the analogy with the Hubbard model, where one-electron excitations in the Mott insulator are frozen out at low
 energies $\omega < \Delta_{MH}$, the Mott-Hubbard gap in the one-electron DOS.  Were one 
to consider the Hubbard model, dynamical {\it bosonic} spin fluctuations, originating from second-order-in-$(t/U)$ virtual 
one-electron hopping processes, would be the natural 
low-energy excitations.  However, in the FKM-like binary alloy model we consider, identifying 
$c\rightarrow c_{\uparrow}, d\rightarrow c_{\downarrow}$ leads to an Ising super-exchange to second order in a 
$(t/U)$ expansion when $U>>t$ in the ``Mott'' insulator.  It is important, exactly as in the Hubbard case, that it is the 
virtual hopping of a $c$-fermion between neighboring sites (from $0$ to $\alpha$ and back
 in our two-site cluster~\cite{ourfirstpaper}) that is necessary to generate such a boson-like mode.  Since this is {\it not} 
a real low energy charge fluctuation, it cannot cause real charge transport.  
But it does lead to a gain $O(-t^{2}/U)$ in super-exchange energy; $i.e$, energy is {\it not} conserved, and so these
 virtual charge fluctuations indeed cause energy transport. 
Physically, this n.n hopping in a gapped ``Mott'' insulator involves creation of a 
particle-hole pair (a holon-doublon composite on neighboring sites).  At low energy, this local ``exciton'' is effectively 
a bosonic mode that disperses on the scale of $J\simeq t^{2}/U$.  These bosons are thus
 {\it not} necessarily linked to any broken symmetry, but naturally emerge in a ``Mott'' insulator.  In our CDMFT, 
the dynamical effects of such ``excitonic'' inter-site correlations on the cluster length scale
 {\it are} fed back into the cluster self-energy, and thus the basic process leading to energy transport but not charge
 transport {\it is} included in CDMFT.  This is also the reason why CDMFT performs much better that single site DMFT when we 
study quantum critical scaling in thermal transport.  The underlying reason for 
this inability of DMFT results to properly describe quantum critical scaling of thermal transport can be understood 
heuristically as follows:  in CDMFT approach, we have argued that thermal transport involves microscopic electronic processes associated with virtual hopping between a given site to its neighbors 
and back.  Such second-order-in-hopping processes block charge transport, but allow energy transport, since such processes 
involve a gain of ``super-exchange'' (of Ising form for the FKM) energy.  In single site DMFT, this process is $O(1/d)$,
 and so is not adequately captured.  But precisely such a process {\it is} captured in our CDMFT, since the dynamical effects 
of inter-site (intracluster) correlations {\it are} fed back into CDMFT self-energies by
construction~\cite{ourfirstpaper}.  These ``bosons'' are thus natural candidates that can account for
 our finding of $K_{el}(T)\simeq T^{1+\nu}$ in the proximity of the MIT.

   Very interestingly, a series of careful experiments on two-dimensional electron gases (2DEGs) show remarakable
features~\cite{narain}: $(i)$ in the low-$n_{s}$ regime where $\rho >> h/e^{2}$, the activated $T$-dependence of $\rho_{dc}(T)$ shows a remarkable ``slowing down'' to an extremely bad metallic state, even as $\rho_{dc}(T\rightarrow 0)\simeq 250h/e^{2}$, $(ii)$ in the {\it same} $n_{s}$-regime, the thermopower shows hugely enhanced values (two orders of magnitude above the Mott value) and, perhaps even more remarkably, exhibits linear-in-$T$ behavior reminiscent of normal metals precisely below $1.0$~K.  It may be possible to apply our high-$D$ approach, which focuses on short-ranged correlations, to these mesoscopic systems {\it if} one could model the system as a 2DEG influenced by strong scattering from atomic-sized (strong) scattering charged centers.  In light of our calculations, the dichotomy between the $T$-dependence of $\rho_{dc}(T)$ and $S_{el}(T)$ can be interpreted as follows: a real charge excitation is blocked in the ``strong-disorder'' limit of the FKM near the MIT due to blocking effects associated with Mottness, explaining the extraordinarly high $\rho_{dc}(T\rightarrow 0)$ below $1.0$~K.  But 
a collective particle-hole (or holon-doublon composite in Hubbard model lore) excitations are real low-energy electronic collective modes that naturally arise in this regime, and lead to a hugely enhanced $S_{el}$.  It is
interesting that our strong-coupling approach seems to rationalize the very unusual experimental observations in a single picture which emphasizes proximity to a (Mott-like) localization transition.  That such observations maybe subtle manifestations of novel phase fluctuation effects is not inconsistent with our view either, since it follows directly from the number-phase uncertainty principle that increasing proximity to electronic localization will necessary generate large phase fluctuation-dominated state(s).

	It is interesting to compare our CDMFT technique of studying thermal transport to the recent work on thermal transport by Finkel'stein and Schwiete~\cite{finkel1,finkel2}. Based on perturbative renormalization group (RG) calculation they studied the quantum criticality using 2+$\epsilon$ expansion and calculate the critical exponent corresponds to different universality classes. This theory describes   the system with both disorder as well as interaction and treat the system as disordered Fermi liquid with disorder induced renormalized Landau parameter. 

Despite the great success of this approach, there are certain limitations - 
(a) In perturbative RG, low temperature excitations are adiabatically connected to non-interacting (but disordered ) electrons. Hence, these excitations which are assumed to be fermionic in nature, play a leading role and collective excitations play a sub-leading role in low temperature region. While in CDMFT approach, fermionic like excitations are absent and the collective excitations play prominent roles. (b)  Perturbative RG is unable to detect any metastable states (like glassy dynamics) arising due to the competition between disorder and interaction whereas our approach can easily capture those features. 

   To summarize, we have showed clear quantum-critical scaling features in $S_{el}(T), K_{el}(T)$ and $L_{0}(T)$ at the MIT strongly testifies to robust quantum critical scaling of thermal transport at a continuous MIT. Ours is a truly microscopic
 approach, and is best valid in the strong localization regime ($k_{F}l\simeq 1$), 
where a Hubbard-like band-splitting type of MIT obtains.  This is the limit opposite to the well-studied weak localization (WL) 
case, where a perturbative-in-$1/k_{F}l$ expansion is possible: at strong 
localization, the criticality is better rationalized in terms of a locator expansion~\cite{sudip1997}, and exhibits signatures 
expected of a continuous ``Mott'' quantum criticality.  Moreover, we are also able 
to connect these critical features in a very transparent way to those observed in electrical conductivity by analysing 
the structure of underlying correlations, thereby providing a direct rationalization for 
our findings.  In view of the fact that the one-band Hubbard model exhibits ``quantum critical'' scaling in $dc$ 
transport near the finite- but low $T$ critical point ($T_c \neq 0$), 
it would also be interesting to study the possibility of related features in thermal transport for such cases in future if the finite-$T$ critical point of the Mott transition could be driven to sufficiently low $T$.

\section*{Acknowledgements}
Authors would like to thank DAE for funding and support.

\end{document}